\documentclass[11pt]{article}
\usepackage{url}
\usepackage{amsmath,amssymb,amsthm}
\usepackage{float}
\usepackage[utf8]{inputenc}
\usepackage{authblk}
\usepackage{multirow}
\usepackage{caption}
\usepackage{framed}
\usepackage{algorithm}
\usepackage{algpseudocode}
\usepackage{enumitem}
\setlist{nolistsep}
\usepackage[compact]{titlesec}

\newtheorem{theorem}{Theorem}[section]

\newcommand{\comment}[1]{}

\usepackage{graphicx}
\usepackage{multicol}
\usepackage{times}
\usepackage{floatflt}
\usepackage[all]{xy}

\usepackage{calc}

\newlength{\myrightmargin}
\newlength{\myleftmargin}
\newlength{\mytopmargin}
\newlength{\mybottommargin}

\newcommand{\ignore}[1]{}

\author{Shridharan Chandramouli }
\affil{University of Utah}

\date{May 3, 2015}
\vspace{2cm}

\title{Parallel Algorithm For Finding The Minimum $s-t$
Cut in a Structured 3-Dimensional Proper Order Graph }

\begin{document}
\begin{center}
Project Advisory Committee:\\[0.5em]
Prof.\ Ross Whitaker (chair)\\
Prof.\ Martin Berzins\\
Prof.\ Suresh Venkatasubramanian
\end{center}
\bibliographystyle{acm}

\vspace{\fill}

\maketitle
\thispagestyle{empty}

\vspace{\fill}

\clearpage
\tableofcontents
\clearpage

\setcounter{page}{1}
\newpage
\pagestyle{plain}

\section{Introduction}
The problem that we aim to solve is that of parallelizing the $s-t$ mincut/maxflow problem for large structured 3D proper ordered graphs in order to improve the performance of current image volume segmentation algorithms. The graph cuts algorithm have been used quite extensively in computer vision and in image processing as a tool for solving energy minimization problems. One of the first uses of graph cuts for computer vision problems was through the work of Greig, Porteous and Seheult where graph cuts were shown to  solve the maximum a posteriori estimation of a binary image in the context of image smoothing by finding the maximum flow through an associated image network.

We are interested in the image segmentation problem viewed as a surface estimation problem applied to segmentation of horizons in geological images. Wu and Chen in \cite{wuchen} studied the image segmentation problem  and gave the basis for solving the energy minimization problem applied to image segmentation as solving the maxflow problem in a multi-column proper order graph. These proper order graphs  are structured graphs arranged as multiple columns in a 3D space with edges from nodes in a column to nodes in its adjacent columns. While there have been many generalized algorithms for solving the maxflow problem, the large size of the image volume and a correspondingly large graph structure often make these algorithms too slow for practical applications. In this project we create a parallelization scheme to solve the maxflow problem for this special graph structure that both improves the running time of these algorithms and reduces the memory footprint of the algorithm by using implicit addressing schemes.

\subsection{Motivation}
Image segmentation is one of the most widely studied  problems that occur frequently in medical imaging and computer vision problems. The use of graph cuts for pixel classification was popularized by Boykov and Jolly in ~\cite{boykovjolly} where they used proposed a new technique for general purpose interactive segmentation of N-dimensional images that gave globally optimal solutions. They posed the problem as an energy minimization problem where the total energy minimized was a sum of the region energy, which represented the relative importance of the region properties, and a boundary term, which was a penalty for a discontinuity in the labeling across neighbors in the image. Wu and Chen extended the idea of image segmentation to surface extraction problems in ~\cite{wuchen}. The energy minimization scheme with respect to the surface $\mathcal{S}$ and the corresponding cost function is given by

\begin{align}
E(\mathcal{S}) = E_{feature}(\mathcal{S}) + E_{prior}(\mathcal{S})
\end{align}

Here $\mathcal{S}$ is defined over a special graph structure whose vertices are arranged in the form of multiple columns. $E_{feature}(\mathcal{S})$ represents the feature matching term that quantifies the desired feature along columns and $E_{prior}(\mathcal{S})$ is the prior term that encodes shape and smoothness information in the form of interaction between columns. Wu and Chen designed a special geometric graph in the form of a multi-column proper-ordered graph. By using such a graph, they proved the equivalence between the energy minimization problem and min-cut on the $s-t$ graph via optimal net surface problems, thus providing a globally optimal and efficient solution. These surface-net problems induce costs on surfaces rather than regions. The graph construction is discussed in detail in the next section.

The min-cut/maxflow algorithms, while having polynomial running time, are still not  scalable with the size of the image for practical applications. One of the applications is the use of this algorithm to segment horizons in seismic image volumes. These volumes typically tend to contain anywhere between 10-100 million nodes, and the current serial algorithm tends to take a few hours to complete. Furthermore, the unstructured implementations often fail to take advantage of the graph structure and leads to a significant memory requirement in the order of a few tens of gigabytes. We therefore require a effective parallel strategy that can both be scalable and uses an implicit addressing scheme that uses considerable lesser memory.

\subsection{Proper Order Multi-Column Graph Structure}
\begin{figure}
\centering
\includegraphics[width=0.4\textheight]{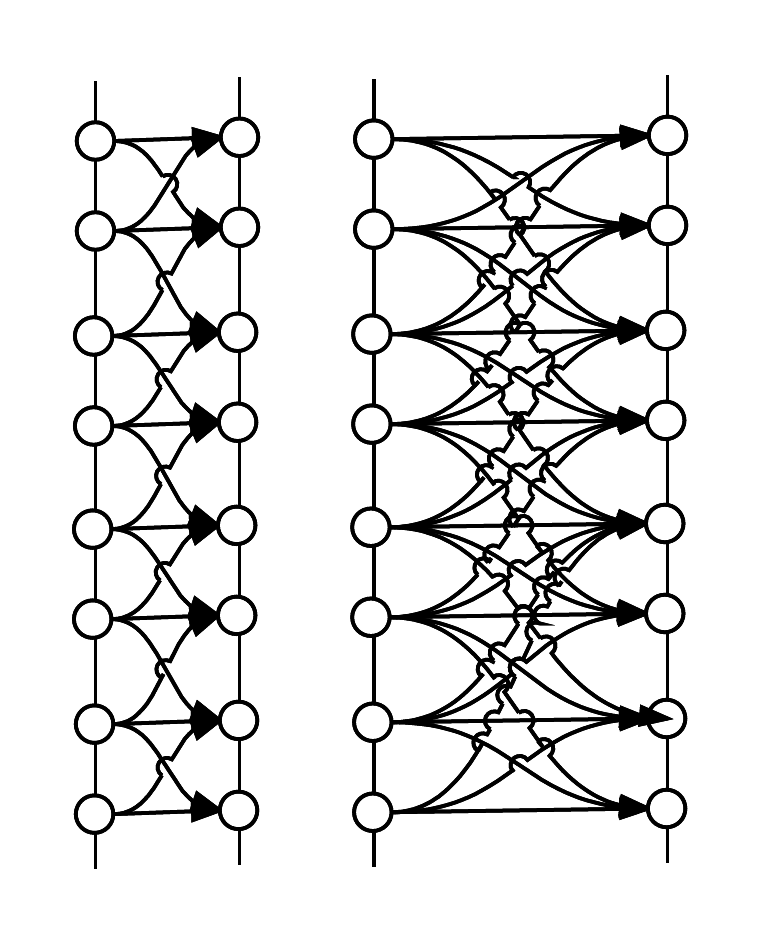}
\caption{\small{Figure shows the proper order edge structure between adjacent columns in the graph. Each node is connected to the immediate node above and below it and a set of vertices in its adjacent column. The number of connections depends on the edge interval. The figure on left has an edge interval of 2 where as the figure on the right has an edge interval of 3.}}
\label{fig:properorder}
\end{figure}

The proper ordered graph is a structured multi-column graph $\mathcal{G} = (\mathcal{V},\mathcal{E})$ with a set of costs on the vertices $\mathcal{V}$ and the edges $\mathcal{E}$. For a multi-column graph, the vertices are arranged logically as a collection of approximately parallel columns, with each column comprising the same number of vertices. A net surface $\mathcal{N}$ is defined by a bijective function in a multi-column graph such that $\mathcal{N}$ intersects with each column at exactly one vertex. A proper order graph also has the following structure. Each column vertex is connected to a sequence of consecutive vertices from each of the adjacent columns. The set of vertices that a column vertex is connected to is called the edge interval of that vertex. For any two consecutive vertices in a column, the first/last edge in the edge interval of a vertex must be above/below or equal to the first/last edge of the vertex below it. For the surface extraction, we are interested in the optimal VCE-weight net surface problem, where each vertex $v \in \mathcal{V}$ in a graph carries a real-valued weight $w(v)$ and each edge $e
\in \mathcal{E}$ holds a real valued cost $c(e)$ while maintaining convexity among the edges of an edge interval. Then the optimal surface is a net $\mathcal{N}$ in $\mathcal{G}$ such that $\sum_{v \in \mathcal{V}(\mathcal{N})} w(v) + \sum_{e\in \mathcal{E}(\mathcal{N})} c(e)$. The graph is then converted to a $s-t$ graph by adding two additional vertices $s$ and $t$ which form the source and sink respectively. The weights on the nodes $w(v)$ are correspondingly converted to the costs associated with the edges connecting the node $v$ to the source or the sink. The figure in ~\ref{fig:properorder} shows the edge structure between adjacent columns in the proper order graph.

\begin{figure}
\centering
\includegraphics[width=0.5\textwidth]{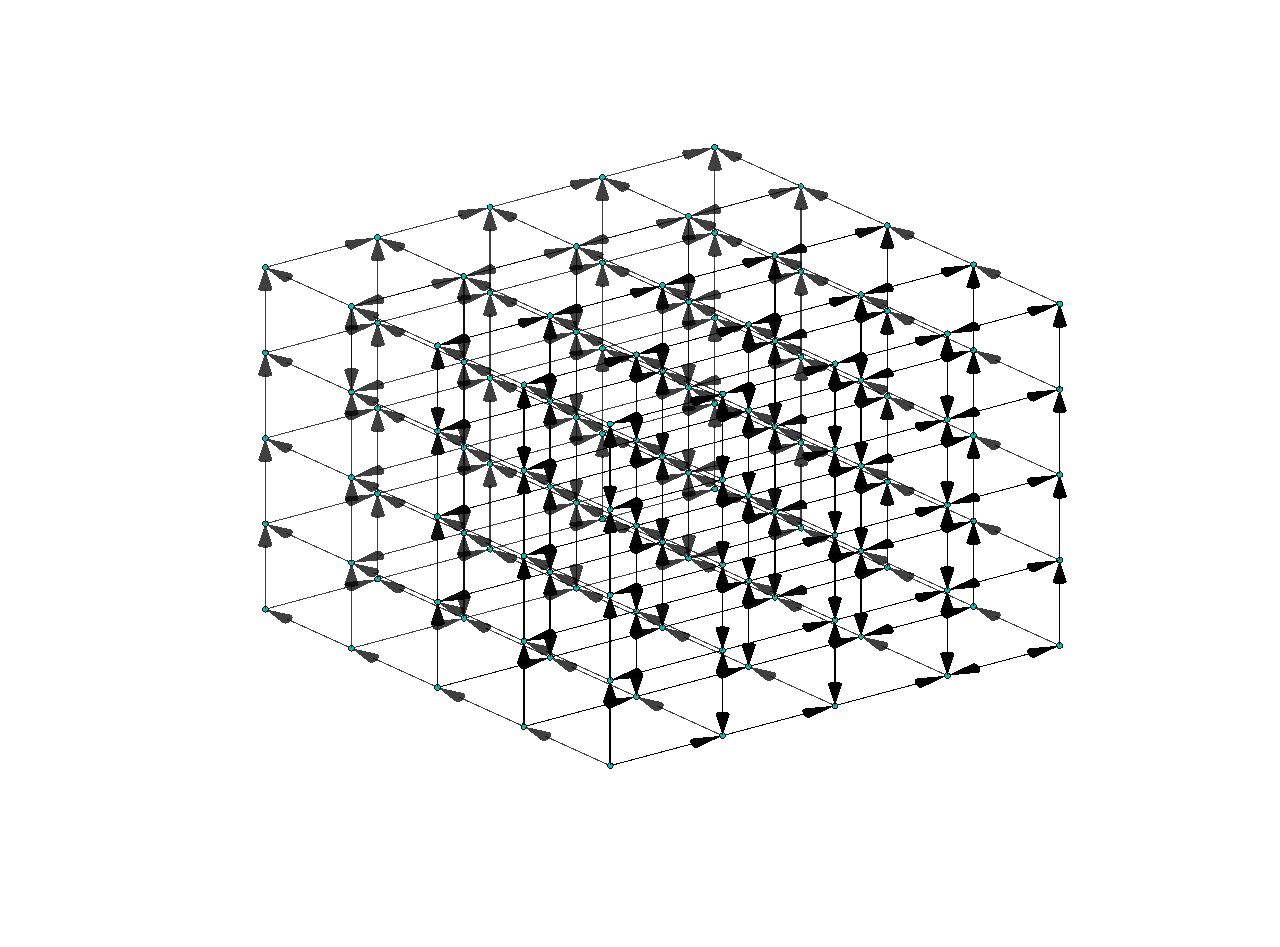}
\caption{\small{A simplified 3D proper order graph structure}}
\end{figure}

\subsection{General Maxflow/Mincut Problem Definition} \label{section:problemdef}
Let  G = (V, E) be a directed connected graph with vertex set V and edge set E. Let $n$ and $m$ be the size of the vertex set $V$ and the edge set $E$ respectively. We call G a flow network if it has two distinguished vertices, a \textit{source s} and a \textit{sink t}, and a positive capacity \textit{c(v,w)} on each directed edge (v,w). A flow \textit{f} on G is a real-valued function on vertex pairs satisfying the conditions below.
\begin{align}
    f(v,w) &\le c(v,w) \hspace{2em} \forall (v,w) \in E\\
    f(v,w) &= -f(w,v)\\
    \sum\limits_uf(u,v) &= 0 \hspace{2em} \forall v \in V - {s,t}
    \label{eq:augconditions}
\end{align}

The antisymmetry constrain (2), used by Sleator~\cite{Sleator:1983:DSD:61337.61338} is for (i) eliminate both edge having positive flow values and (ii) simplifies the formal expressions of constraints.

The value of a flow \textit{f} is the net flow into the sink,
    \begin{equation}
    |f| = \sum\nolimits_{v \in V} f(v,t)
    \end{equation}
    A maximum flow is a flow of maximum value. The problem which we solve is that of computing a maximum flow in a given network.

    A \textit{cut} S, $\bar S$ is a    partition of the vertex set $( S \cup \bar S\ = \ V, S \cap \bar S \ = \    0)$ with $ s \in S$ and $t \in \bar S$. That is, a cut is a dichotomy of the input vertices in such a way that the source and the sink vertices $s$ and $t$ do not belong to the same set. The capacity of the cut is
    \begin{equation}
    c(S, \bar S) = \sum\nolimits_{v \in S, w \in \bar S} c(v,w)
    \end{equation}
	The flow across the cut is
    \begin{equation}
    f(S, \bar S) = \sum\nolimits_{v \in S, w \in \bar S} f(v,w) = |f|
    \end{equation}

\begin{theorem}
For any network $G$, the value of a maximum $s-t$ flow is equal to the capacity of the minimum
$s-t$ cut.
\label{theorem:ford}
\end{theorem}
Fold Fulkerson in \cite{fordfulkerson} had proved that the problem of finding the maximum flow is the dual to the problem of finding the minimum cut.

\pagestyle{plain}

\section{Related Work}
\label{sec:relwork}
\par
The maxflow/mincut algorithms were first stated by US Air Force researcher Theodore E.Harris and retired army general Frank S.Ross, for determining the most efficient way of disconnecting the railway lines in the Soviet Union. Much of the earlier solvers for the maxflow problems were algorithms that used linear programming to solve the problem. Ford and Fulkerson in their paper first gave an efficient  augmenting paths algorithm to solve this problem which essentially worked by repeatedly finding a "minimal path", that is a path from the source to the sink, and finding the edge with the minimal or "bottleneck" capacity. This version of the augmenting paths algorithm had a running time complexity in the order of $O(mf)$ where $m$ is the number of edges in the graph and $f$ is the maximum flow through the graph. This is because a path from $s-t$ could be found in $m$ time and each time a path is found, the flow is increased by a value of atleast $1$, giving an upper bound of $O(mf)$. Edmonds-Karp~\cite{Edmonds:1972:TIA:321694.321699} provided an implementation of the Ford-Fulkerson algorithm which defined the search order for finding the augmenting paths in the network or graph.  At each stage, Edmonds-Karp algorithm tried to find the shortest $s-t$ path in the network by doing a BFS on the graph. It could be seen that at every stage of the algorithm, the length of the path found increases monotonically, and therefore finding a path requires $O(m)$ time, augmenting the path requires $O(n)$ time and finally updating the residual graph takes $O(m)$ time, giving a $O(nm^2)$ algorithm. Dinitz~\cite{dinic} in 1970, provided a "blocking flow" algorithm which improved the bounds to $O(n^2m)$. Dinitz algorithm, used a concept of level graphs where each vertex was assigned a label that is equal to the distance of that node from the source. At each stage of the algorithm, the level graph was constructed and only the edges with capacity that transferred flow to the vertex with a higher label were added to the residual graph. Once all such paths are exhausted, a new BFS is done to update the labels and the residual graph is updated again. This process is repeated until there are no paths that can carry flow to the sink. In other words, the Dinitz's algorithm tries to find all the paths from the source to sink, which have the same length and tries to saturate the maximum possible edges from these paths.

The notion of preflow was first introduced by Karzanov~\cite{karzanov} in 1974. The preflow is a function on the edges that may violate, in a certain way, the flow conservation condition in nonterminal nodes.%
Karzanov's preflow algorithm was in a way similar to that of Dinitz's algorithm in that the algorithm consisted of $O(n)$ stages, each of which solved a smaller auxiliary problem of finding the blocking flow in a layered network. As in the Dinitz's level graphs, each of the vertices in Karzanov's algorithm maintained its distance to the sink and flow was between successive layes in this graph. However, instead this algorithm introduced and used a couple of functions. The first is the $excess$ of a node. The excess value of a node is the difference of the flow coming into the vertex and the amount of flow going out of the vertex is to be remembered, that by the flow conservation requirement of the problem definition, this value is equal to zero in all the augmenting paths algorithms discussed thus far.

More formally, for any vertex in the graph $v \in V$ let $\delta^{in}(v)$ be the set of vertices that bring in flow to this vertex and let $\delta^{out}(v)$ be the set of edges that take flow away from this vertex. Given a flow function $f$ as defined in section \S\ref{section:problemdef}, the excess of $f$ at a node $v \in V$ is defined as below.

\begin{align}
e_f(v) = \sum_{(u,v) \in \delta^{in}(v)}f(u,v) - \sum_{(v,w) \in \delta^{out}(v)}f(v,w)
\end{align}

If for some function $f^\prime : E \to \mathbb{R}_+$, $e_{f^\prime}(v) \ge 0$, Karzanov called such a function a $preflow$. It is to be noted that the preflow $f^\prime$ is still lesser than the capacity function for every edge $(u,v) \in E$. A flow or preflow $f$ is called blocking if any directed path from $s$ to $t$ contains at least one $saturated$ edge $e = (u,v)$, that is the flow $f(e) = c(e)$.

All the vertices in $G$ are ordered toplogically, that is every node is labelled in such a way that a node that is closer to the sink has a higher label, than one that is farther away(ties are broken arbitrarily). Doing this requires a breadth first search which would take $O(m)$ time.

The algorithm works by repeating a set of operations called the ``push'' and ``balance'' operations. During a $push$ operation, the algorithm tries to push flow from a node with a lower index in the topologically sorted order a node with a higher index. In otherwords, the push operation simply moves flow towards the sink. The balancing operation in effect tries to return the excess flow which cannot be pushed forward, back to the sink. The number of push operations that could be performed is $O(n)$  for each vertex and since there are only $n$ vertices in the graph, the bounds for the push operation is $O(n^2)$ This is the same as the number of operations for the total numbe of balancing operations. Therefore the complexity of finding the blocking flow is bounded by $O(n)$. For each blocking flow that is found, at least one vertex gets disconnected from the sink, therefore the total running time of the algorithm is bounded to $O(n^3)$.

In the next section, we have discussed two algorithms viz. the push relabel algorithm and the Boykov Kolmonogrov's modified augmented path algorithms, in detail.

\pagestyle{plain}

\subsection{The Push-Relabel Algorithm} \label{sec:pushalgo}
The push relabel algorithm works on the same notion of $excess$ as Karzanov but finds the maximum flow in a graph by repeatedly performing two basic operations called the push and relabel operations. While Karzanov's preflow algorithm used a similarly termed operations, it was solely for finding a blocking flow in the graph. In the Push-Relabel algorithm however, these two operations are guaranteed to converge and produce a flow value that is maximal.

In order to move the flow excess from one vertex to another and in order to estimate the distance from a vertex to the sink or the source, Golberg and Tarjan introduced two additional functions on the vertices of the graph. The concept of residual capacity (which is in effect the same as the capacity of the edges in the residual graph for the augmented paths algorithm) is used to determine the difference in the capacity of the input edge and the flow going through that edge at any point in the algorithm. That is $r_f(u,v) = c(u,v) - f(u,v)$ The terms saturated and unsaturated edges are similarly borrowed to determine if the residual capacity is zero or positive respectively. Any edge is called a residual edge if $r_f(u,v) > 0$. In order to estimate the distance of a vertex (either the sink or the source), a new function called a $valid labelling$ is used. The $valid  labelling$, $0 \le d(v) < \infty$ gives a lower bound on the estimated distance from that vertex to the sink .  Initially the values for the source and sink are set as $d(t) = 0$, $d(s)= n$ and finally for every residual edge between $(u,v)$, the following condition holds true $d(v) \le d(w) + 1$ .  Since the sink has the lowest possible label $d(t) = 0$, if $d(v) \ge n$, then $d(v) -n$ gives the lower bound on the actual distance to s in the residual graph.  Finally any vertex in the graph, which has a distance label $d(v) < \infty$ and an excess of $e(v) > 0$ is called an active vertex. The algorithm works by repeating the following two basic operations until they are applicable to any of the vertex in the graph.

\textbf{Push Operation} The push operation is in essence very similar to the definition proposed by Karzanov, in that it attempts to push the excess towards the sink (or to the source in case it cannot reach the sink. Explanation below). The push operation operation for two nodes $(u,v)$ tries to move the maximum possible preflow from $u$ to $v$ when $u$ is active, $r_f(u,v) > 0$ and $d(v) = d(w)+1$.  After the push operation is completed, the flow $f(u,v)  = f(u,v) + \delta$, $f(w,v) = f(w,v) - \delta$ and the excess $e(u) = e(u) -\delta$, $e(v) = e(v) + \delta$. A push operation can make the node $v$ active, if the previous excess was $0$.

\textbf{Relabel Operation} The relabel operation is applicable to any $v$ which is active and $\forall w \in V$, $r_f(v,w) > 0 \implies d(v) \le d(w)$. In otherwords, the relabel operation is applicable to all the nodes which cannot push flow to its neighbors because the neighbor is not neccesarily closer to the sink in the graph that it currently is. A relabelling of $v$ sets the label of $v$ to the largest value allowed by the valid labeling constraints. It does this by scanning all the residual edges in the graph going out of $v$ and updating $d(v) = min {\{ d(w)  + 1 | (v,w) \in E_f \}}$.

The algorithm works as follows. Initially for all the outgoing edges $(s,u)$ from the sink the flow along  $f(s,u) = c(s,u)$ and $e(u) = c(s,u)$ and  the reverse flow $f(u,s) = -c(s,u)$. All the vertices in the graph as labelled as $d(s) = n,  d(t)  = 0 ; d(V-\{s,t\}) = 0$.\footnote{It is to be noted that a better labelling strategy would be to assign the real distance to the sink by performing a reverse BFS from the sink. However, the bounds derived using this labelling is valid both these approaches}. All the vertices with preflow marked as active and the push and relabel operations are repeated consecutively until neither the push and the relabel operations are not applicable.

There are generally two kinds of optimization heuristics that help
speed up this algorithm and they are the "Relabeling" and the
"Optimal selection" heuristics. Both these strategies were studied
and  popularized by Cherkassky~\cite{cherkassky}.
Global relabeling is a way to enable the algorithm to find the most
straight forward path to the sink in order to push flow through it.
The $height$ of $label$ of a vertex, though represents the distance
to the sink in the level graph of the initial residual graph, the
continuous push and relabel operations can effectively change their
labels to be completely different from their actual distance. This is
because of the fact that the ordering of the capacity carrying edges
from a vertex plays a role in way that the flow is pushed through the
graph. Since the label of a vertex effectively states upper bound on
the distance to the sink after the first few iterations of the
algorithm, such an upper bound may become less and less tight as the
algorithm proceeds. This is also in effect the lack of the global
picture when performing the push and relabeling operations as they
work only withing their neighborhoods. Cherkassky~\cite{cherkassky},
ensures that the label on the vertices are brought back to their true
values by performing a global breadth first search once in a while.
This idea was called the global relabeling, as we have a global
picture of the distances to the sink and "know" which of the paths
are better to reach the sink faster. The general idea here is that
after performing a fixed number of discharge operations, the Push-
Relabel algorithm is stopped and the labels are updated as descibed
above using a BFS. The number of iterations is generally chosen to be
as a constant factor of the number of nodes in the graph, with the
factor ranging from $1 - 2$. While this approach increases the bound
on the algorithm by $O(m)$, in all practical implementations, this
approach has always reduced the running time of the Push-Relabel
algorithms significantly.

Another relabeling technique, developed
independently by Cherkassky and Derigs and
Meier~\cite{Cherkassy:1995:IPM:645586.659457} was the concept of the
Gap relabeling. This is based on the following observation
\footnote{Notations and definition borrowed from
Cherkassky~\cite{cherkassky}}. Let $g$ be an integer and $0 < g < n$.
Suppose at certain stage of the algorithm then there are no nodes
with the distance label $g$ but there are nodes $V$ with $g <
label(v) < n$. Then this sink is not reachable from any of these
nodes. Therefore, the labels of such nodes may be increased to $n$.
Cherkassky used a linked list of buckets, one for each of the
possible labels $1 .. n$ to determine if there were no active
vertices of that particular label. As soon as an empty bucket was
found, all the contents of the buckets with labels greater than this
was moved to the last bucket with label $n$.

The optimal selection strategy defines the order in which the vertices are processed. The two most commonly used strategies are the First In First Out (FIFO) where the vertices are processed in the same order that they became active and the highest label processing where the vertices with the highest label was processed first.

Cherkassky in his survey of Heuristics, found that the Highest label
with gap and global relabeling performed (HIPR) the best for almost
all types of graphs, where as the FIFO selection rule with Gap and
Global relabeling(FPR) was the second best. In the absence of
the gap relabeling, Highest Label in fact performed poorly in
comparison with FIFO and was the slowest among the classes of
heuristics analyzed by Cherkassky.

\pagestyle{plain}

\subsection{The Boykov-Kolmogorov Algorithm}
In ~\cite{Boykov:2004:ECM:1018034.1018355} Boykov and Kolmogorov gave a modified version of the augmented paths algorithm of Dinic's for the vision problem. The algorithm essentially works by maintaining two search trees one from the source and one from the sink. The nodes are labeled as either active, passive or free, where the nodes at the frontier are the active nodes, while the nodes that are present as part of the search tree but not at the frontier are called the passive nodes. All other nodes are free. In the "growth stage", both the source and the sink search trees frontiers are expanded by one level by doing a breadth first search along capacity carrying edges. Whenever the two search trees meet, it indicates a path from the source to the sink that is of the shortest length. The bottleneck flow, when passed through the path in the "Augmentation stage", could make some of the nodes in the search trees as "orphans". That is, the path linking these nodes to their parent node in the search tree gets saturated and is therefore not a valid part of the search trees. In fact, the augmentation phase may split the search trees into forests. Boykov and Komogorov introduced an "adoption" stage to restore the single tree structure of the two search trees with roots in the sink and the source. For every node that was "orphaned", a new parent is found by looking at all the node's neighbors that can carry flow and is connected to either the source search tree or the sink search tree. If such a node is found, then the orphaned node's parent is updated as its neighbor and the next orphaned node is processed. If no such node could be found, then the orphaned node and all its former children are  marked as "free". The adoption stage completes when there are no more orphaned nodes. The algorithm loops through the growth, augmentation and adoption stages until no more work could be performed.

Boykov and Kolmogorov showed that the above algorithm, while having a complexity of $O(mn^2|C|)$ outperformed all other standard algorithms.

\pagestyle{plain}

\subsection{General Parallel Maxflow Algorithms}
The first parallel algorithm that was based on Dinic's blocking flow algorithm was given by Shiloach and Vishkin in \cite{Shiloach:1982:ONL:61335.61336}. Though a number of attempts were made at parallization, it was generally hard to achieve a good performance boost with the parallel version of the augmenting paths algorithm due to the fact that there has to be a global picture of the entire graph in order to find an augmenting path.
Golberg and Tarjan in ~\cite{Goldberg:1986:NAM:12130.12144} gave a parallel algorithm for their Push Relabel algorithm that ran in $O(n^2logn)$ time with $n$ processors and $O(m)$ additional space. The first practical implementation of the push relabel algorithm was given by Anderson and Setubal in ~\cite{Anderson:1995:PIP:213123.213125}. They introduced the concepts of "Waves" that helped perform global relabeling in parallel with the push and relabel operations. Our approach to parallelization, as will be seen in the next sections, borrows the ideas from this paper and applies it to the structured graph setting. They imposed a function on the vertices called the ``wave number'' which enabled them to perform concurrent global relabelling. The wave number of a vertex was initially  set to zero and everytime a global relabelling was performed on the vertex, the wave number was increased. This function helped impose the condition that labels are always consistent for vertices that belong to the same wave number. The FIFO queue was converted into a hierarchical set of queues, where each processor had a local inqueue and outqueue and there was one global shared queue. Every time a processor needed to process vertices, it picked a constant number of vertices from the global queue and stored them in the local inqueue. Any time the processor needs to put a vertex into the queue, it did this by placing the vertices in the local outqueue and moving them to the global shared queue only when the outqueue gets filled. This kind of processing helped reduce the amount of
false sharing between shared memory systems and generally improved the performance of the algorithm. All the pushes and relabelling operations required locks on the respective vertices. Every processor advanced the global relabelling wave once after every $j$ discharges were performed.

In ~\cite{bader} Bader and Sachdeva improved on the idea of Anderson and Setubal and implemented a cache aware version of the push relabel algorithm that used both the "waves" from Anderson and Setubal and a gab relabeling heuristic and highest label processing strategy. While they showed a significant improvement over Anderson and Setubal, we chose not to implement the highest label and gap relabeling strategy due to the uniform partitioning strategy discussed in the following section.

In ~\cite{cudacuts} Vineet and Narayanan implemented the first parallel version of the push relabel algorithm for the GP-GPUs. While the performance improvements through the use of GPUs can be significantly larger than shared memory parallelism, they are constrained by the limited global memory sizes and is impractical for large graphs such as those from our application.

In ~\cite{liu} Liu and Sun provided an parallel version of the BK algorithm that split the structured image volume into segments and used adaptive bottom up merging after processing the smaller segments in parallel. We first implemented this approach to parallelization for the seismic image volumes and found that the performance improvements capped at a meager 3x. This is because, the algorithm inherently tries to exhaust the smaller paths in the graph in parallel and then exhausts larger paths that span segment boundaries after merging the segments. Such an approach still leaves a significant portion of the paths to be augmented in the later stages of the algorithm, therefore limiting the amount of work that could be done in parallel.

\pagestyle{plain}

\section{Parallel Maxflow algorithms for Proper Order Graphs }

We evaluated two strategies for parallelizing the maxflow algorithm. In our first approach, we adapted Liu and Sun's parallel BK algorithm to 3D proper order graphs and in the second approach, we implemented a modified version of Anderson and Setubal's parallel push relabel algorithm.

\subsection{Parallel BK algorithm with hierarchical merging}

\begin{figure}[ht]
\centering
\includegraphics[width=0.5\textwidth]{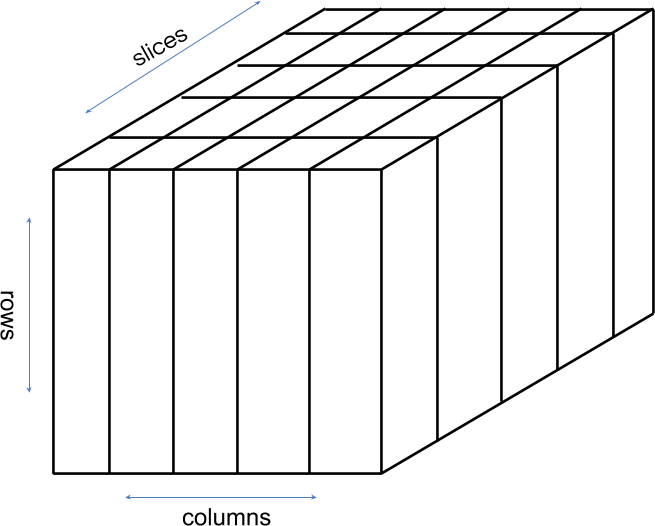}
\caption{\small{Image shows the 3D image volume segmentation strategy for use with the parallel BK algorithm. The segments are partitioned along the columns and slices}}
\label{fig:bkseg}

\end{figure}

\begin{figure}[ht]
\centering
\includegraphics[width=0.5\textwidth]{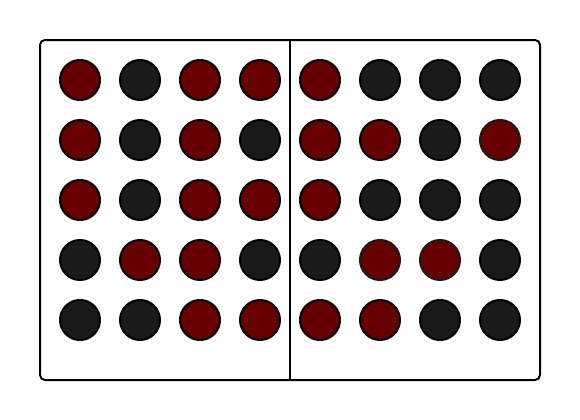}
\caption{\small{The red nodes are connected to sink and the black nodes are connected to the source. After merging, any new path found from source to sink has to include the nodes that are adjacent to one another across old segment boundaries and are marked as belong to different terminal nodes.}}
\label{fig:bkmerge}

\end{figure}

We extended Liu and Sun's parallel BK algorithm to 3D by first segmenting the structured graph into smaller segments of uniform size. The smaller segments were created by splitting the large 3D volume by splitting across the two most dominant dimensions (mostly the rows and columns). The segmentation strategy is shown in the figure ~\ref{fig:bkseg}. Typically the use of narrow band factor limits the number of rows in the volume to be be considerably smaller than the number of columns or slices and therefore the segmentation strategy split the larger volume along the columns and slices as shown in the figure. The number of nodes in each segment was roughly made equal. The algorithm works as follows. In the first stage of the algorithm, the source and the sink search trees are expanded to exhaust all the paths within the small segments. When no further paths are found in the small segments, they are hierarchically merged into larger segments by combining adjacent segments into larger segments containing twice the number of nodes as the smaller segments. The sink and source search trees that were constructed in the smaller segments are then further expanded to cover nodes/paths that span the smaller segment boundaries. This way, the paths of larger lengths are found successively till the final stage where the algorithm is run on the entire graph to exhaust all the paths that were not covered in the earlier stages. It is to be noted that when two segments are merged to form a larger segment, any new capacity carrying path has to include the nodes along the segment boundaries as shown in the figure ~\ref{fig:bkmerge}.
Since the segments are logically separated and the paths are restricted to these segments, we do not use any kind of locking in this approach. Also since the segments are generally small, there is a good amount of memory locality that is helpful when processing very large graphs in a shared memory system.

\subsection{Parallel Push-Relabel algorithm with Level Synchronized Global Relabeling}
One of the major issues with the BK algorithm was its scalability or the lack thereof. This is largely due to the fact that the augmented path algorithms require a global picture of the entire graph and the segmentation strategies used for parallelization often make this requirement difficult to satisfy. The push-relabel algorithm however, works locally within its immediate neighborhood and are therefore better candidates for parallelization. In order to parallelize the push relabel algorithm, we adapted the concept of "waves" from Anderson and Setubal and introduce a level-synchronized global relabelling strategy that helps perform global relabeling and the push-relabel operations in parallel in a shared memory system with limited communication between the segments.

\subsubsection{Segmentation strategy and vertex processing}
Both \cite{Anderson:1995:PIP:213123.213125} and \cite{bader} used a hierarchical queue or bucket structure where there was a local queue/bucket for every processor and a global queue that was common to all the processors. The nodes in the graph did not have any processor affinity and every time a processor's local queue became empty, it received a fixed set of nodes from the global queue to start processing. The movement of vertices between the global and the local queues were done in batches so that there is limited locking of vertices. While this strategy works for a generalized graph, with our proper ordered graph, we decided to assign processor affinity to the nodes.

\begin{figure}[ht!]
\centering
\includegraphics[width=0.6\textwidth]{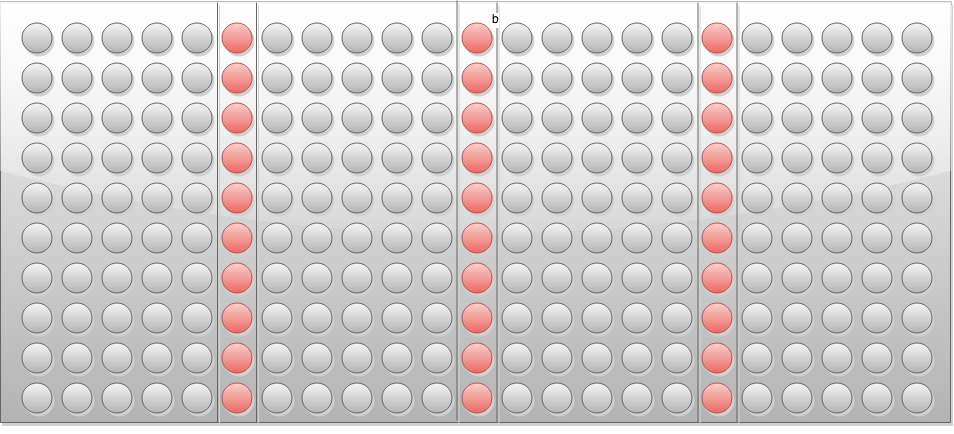}
\caption{\small{The orange nodes are shared between adjacent segments and needs to be locked before processing. The gray nodes are interior to the segment and need to be locked only when interacting with a shared vertex.}}
\label{fig:segmentstrategy}
\end{figure}

Depending on the number of processors, the proper order graph was divided into a number of smaller segments along the columns. In almost all of the image volumes , the number of columns far exceeds the number of rows (which is limited by the narrow band factor) and the number of slices in the volume. Therefore, dividing the larger volumes in to $n$ smaller segments along the columns gave us a simple and effective partitioning strategy. These segments, numbered $1 - n$  were each assigned to a separate processor and the vertex affinity was set to that particular processor. Since these vertices are processed by the same processor, the nodes that are interior to the segment do not require any kind of locking mechanism.  There can however, be a transfer of excess flow between adjacent columns that belong to different (and adjacent) segments. Such nodes are called the "shared" vertices/nodes. Any changes that are made to the shared vertex's excess are made in a critical section guarded by a lock. Similarly, any change in the residual capacity of edges that originate from or go to a shared vertex are made after obtaining a lock. The segmentation strategy is shown in the figure ~\ref{fig:segmentstrategy}.

\begin{figure}[ht!]
\centering
\includegraphics[width=0.9\textwidth]{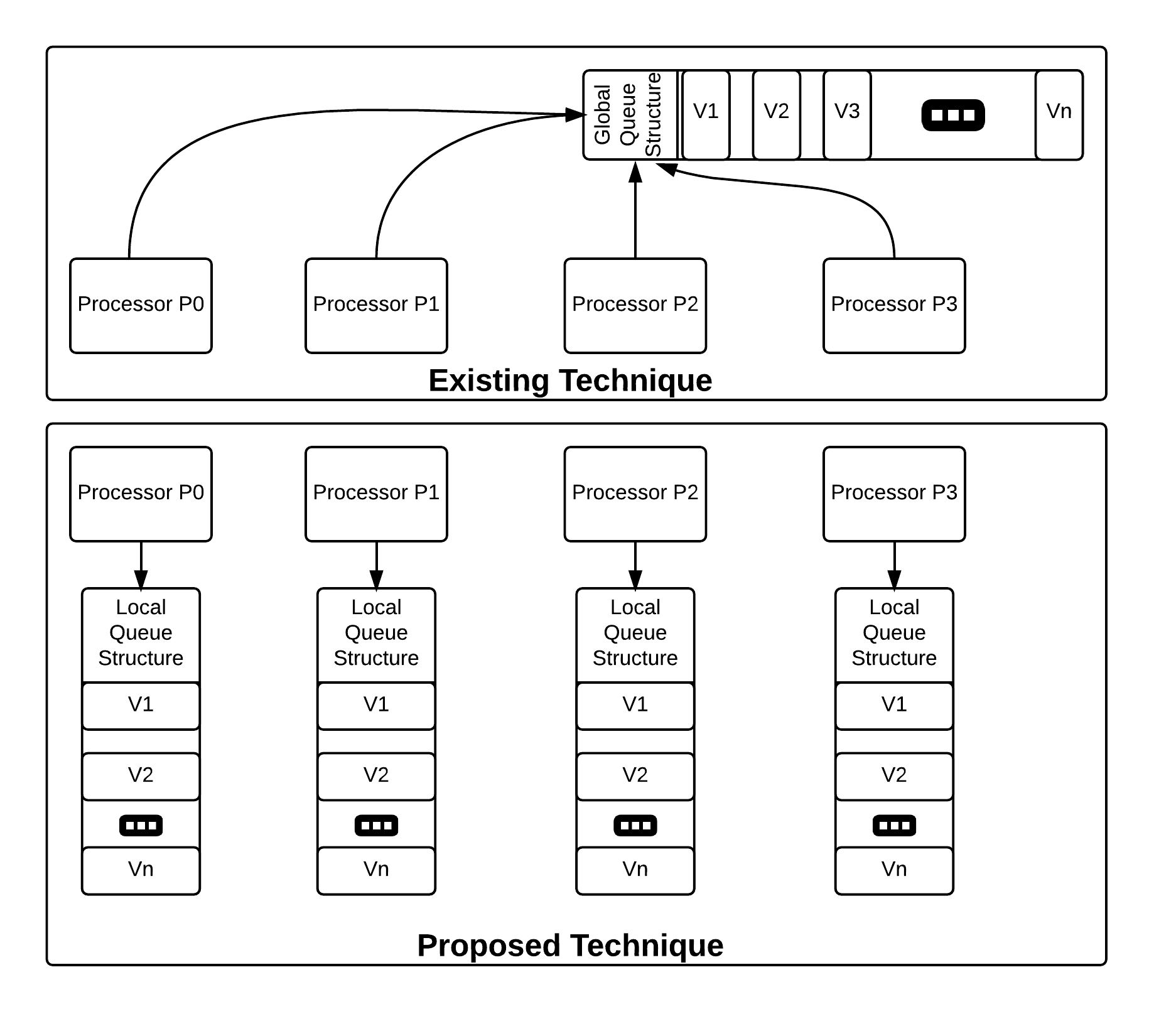}
\caption{\small{Figure shows the existing and proposed vertex processing strategy. Each processor has a local processing queue instead of a shared global FIFO queue.}}
\label{fig:poolstructure}
\end{figure}

With the segmentation strategy and the vertex affinity described above, two aspects become clear. One, setting the vertex affinity meant that we did not require a hierarchical global queue structure. Instead we can replace it with one local queue for every processor. Second, the proper structuring and the segmentation strategy means, any vertex addition to the queue is done locally or from the immediate adjacent segments only. The processors process the vertices in the queue in an local FIFO ordering. The figure ~\ref{fig:poolstructure} shows the queue structure.

\subsubsection{Level Synchronized Global Relabeling}

Anderson and Setubal in~\cite{Anderson:1995:PIP:213123.213125} first introduced the concept of waves to help perform parallel global relabeling concurrently with the push and relabel operations. We use the concept of waves and introduce the concept of level synchronized global relabelling to effectively update the labels of the vertices periodically to reflect the original distance to the sink. The global relabeling in the sequential setting is performed as a reverse Breadth First Search (BFS) and updating their labels to the distance to the sink along edges that can send flow to the sink.

The wave global relabeling works as follows. Every global relabeling from the sink is considered a consecutively numbered "wave" and each vertex has an additional function associated with it, called the wave number. Every time a new global relabeling wave visits a node, its wave number is to set to the global relabeling wave's number. Anderson and Setubal further enforced some restrictions on the push and relabel operations. Any wave that visits a node updates the label only if the new label is greater than the existing label. This ensures that the distance measure does not decrease as the algorithm progresses. Also the push operation is modified to ensure that the flow is only pushed along vertices that have the same wave number. Both these conditions were shown to satisfy the valid labeling criteria of Goldberg and Tarjan.

In our implementation, we used the concept of waves to perform the push and relabel operations in parallel. However, the global relabeling wave front propagation was performed in a level synchronized fashion. Every processor along with a local process queue, also a global relabeling queue. Once a certain number of discharge operations are completed, a new global relabeling wave is started. Every processor in parallel adds all the vertices that are directly connected to the sink through capacity carrying edges to a queue. This effectively advances the frontier by one level. Once all the processors complete advancing the frontier by one level, a software barrier is reached before advancing to the next level. Since the frontier advancement is synchronized at every level, we name this way of global relabeling as a Level Synchronized global relabeling scheme. A software barrier is required at each level of the frontier advancement because a shared vertex could be reached by the same wave at different times and different levels from the two adjacent segments, which can lead to the wrong label being assigned to the node.

\begin{figure}
\centering
\includegraphics[width=0.4\textwidth]{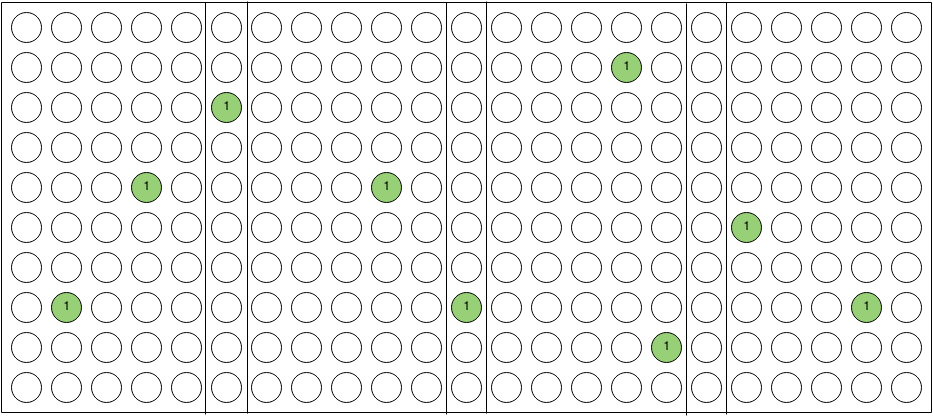}\\
\includegraphics[width=0.4\textwidth]{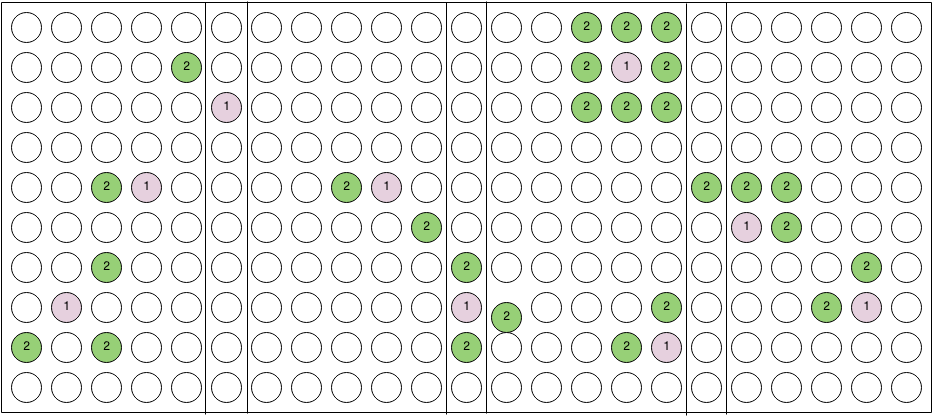}\\
\includegraphics[width=0.4\textwidth]{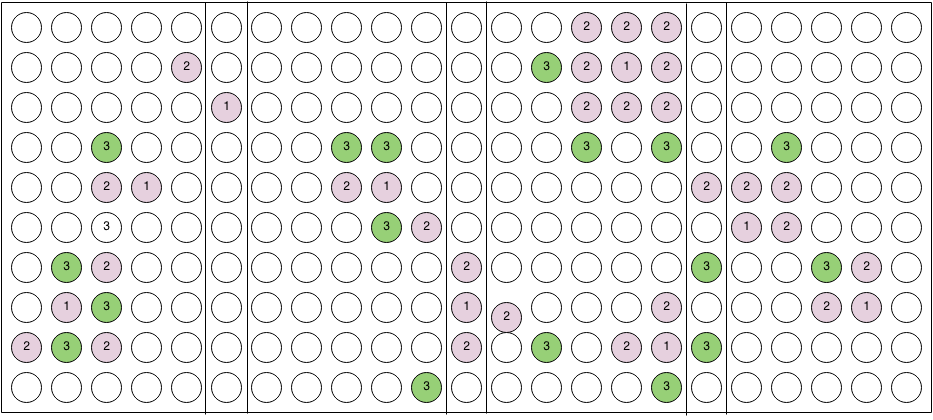}
\caption{\small{Figure shows the frontier expansion during the level synchronized global relabelling. The green vertices represents the vertices at the frontier and the pink vertices are those which have been relabeled in this current wave. All other nodes can become part of this wave as the wave progresses. }}
\end{figure}

\subsubsection{Structured Graph and Implicit Addressing Scheme}
With large graphs, we needed an implicit addressing scheme that would reduce the memory requirements of the algorithm. In our initial implementation, the edges were maintained as a linked list associated with a node. Each edge had an associated pointer that pointed it to the mate edge and another pointer to the node connecting this edge apart from the residual capacity. With the structured graph, we can know beforehand the number of edges that are required for a node and we create contiguous blocks of edge capacities where a node's starting edge can be known by multiplying the node index with the number of edges per node. The edges are arranged such that the offset within a node's set of edges indicates the direction of the edge from that node. Since we know the number of slices, rows and columns before hand, the edge mate can be determined using the direction offset and the node's index. We also cached the results of offset calculation in a 3D matrix (size: rows $\times$ slices $\times$ edges per node) to reduce the amount of time required for computing the offsets. The size of the cache is significantly lesser than having a mate stored as part of the edge by a factor equal to the number of columns in the image. Since the number of columns is usually the most dominant dimension in the image, this corresponds to a significant reduction in the amount of memory required without a loss in computational performance.

\subsubsection{Termination}

Since each processor works individually without a shared queue, a thread can end only when all the processors finish exhausting all the nodes in their individual  queues. This is because, a single process which has an empty queue at some point may receive flow from neighboring segments at a later time. The process of finding if all the processors have empty queues, however, can become a very expensive operation as a full locking of all the processors is required to ensure correctness. The algorithm in ~\ref{alg:termination} shows the termination process.

To eliminate unnecessary locks, each processor has an associated  \textit{emptyqueue} boolean flag, that can be read by other processors, which indicates if its queue is empty. The boolean flag is set to true when the queue becomes empty and is set to false when an empty queue gets elements added to it. The updates to the \textit{emptyqueue} flag are performed with a global writer lock on a shared mutex that is common to all the processors. There is also a shared boolean flag \textit{terminateflag} which is common to all the processors. The \textit{terminateflag} is initially set to false when the algorithm begins. When a processor finds that its local queue is empty after processing its last active vertex, it sets its $empty queue$ flag to true and checks if all other processors have completed processing by reading  the $empty queue$ flags of all processors after obtaining a global reader lock on the shared mutex. If all other processors have their $empty queue$ flag set, then this processor sets the $terminate flag$ to be true and exits. If on the otherhand, some other processor is still working on its queue, then this processors repeatedly polls the $terminate flag$ till the terminate flag becomes true or this process has active vertices in its queue. Since the updates to terminate flag are not done with any kind of locking, the polling operation does not stop other processors from processing its active vertices.

\begin{algorithm}

\caption{Vertex discharge loop}
\label{alg:termination}
\begin{algorithmic}
\State $dischargecounter \gets 0$
\Procedure{PopQueue}{pid}
\State $v \gets head(queue)$
\If {v is lockable}
	\State lock v.mutex
\EndIf
\EndProcedure
\State
\Procedure{isQueueEmpty}{}
\State Get reader lock on shared mutex.
\For {$i \in 1 .. num\_proc$}
\If {$emptyqueue(i) \ne true$}
\State {return false}
\EndIf
\EndFor
\State $terminateflag \gets true$;
\State return true
\EndProcedure
\State
\Procedure{discharge}{pid}
\State $global\_check \gets true$
\While {true}
\While {Local queue has vertices}
\State $v \gets PopQueue(pid)$
\State $Apply(v)$
\State $global\_check \gets true$
\State $dischargecounter \gets dischargecounter + 1$
\If {global relabel flag set}
	\State $GlobalRelabel(pid)$
\EndIf
\EndWhile
\If {$terminateflag \ne true$}

\If {global\_check = true \& isQueueEmpty() = true}
	\State return;
\EndIf
\State $global\_check = false$
\State yield to other threads
\If {global relabel flag set}
	\State $GlobalRelabel(pid)$
\EndIf
\Else
	\State return
\EndIf
\EndWhile
\EndProcedure
\end{algorithmic}
\end{algorithm}

\begin{algorithm}
\caption{Vertex Process}
\label{alg:vertexprocess}
\begin{algorithmic}
\Procedure {Apply}{Vertex v}
\State let $e = \{v,w\}$ be the current edge of $v$
\While {$e$ is not the last edge}
\If {$e$ can carry flow}
\If {$w$ is lockable}
\State try to lock {$w$}
\If {lock unsuccessful}
	\State set current edge of $v = e$
    \State add $v$ to local queue
    \State \Return
\EndIf
\EndIf
\If {$e$ is admissible}
\State push ($v, w$)
\If {$w$ becomes active}
\State add $w$ to appropriate process queue
\EndIf
\EndIf
\EndIf
\State set $e$ to be the next edge going out of $v$
\EndWhile
\State relabel($v$)
\If {$v$ is still active}
	\State add $v$ to local queue.
\EndIf
\EndProcedure
\State
\Procedure{Push}{$e = \{v,w\}$}
\State cap $\gets$ min\{excess(v), capacity(e)\}
\State m $\gets$ mate(e)
\State capacity(m) $\gets$ capacity(m) + cap
\State capacity(e) $\gets$ capacity(e) - cap
\State excess(v) $\gets$ excess(v) - cap
\State excess(w) $\gets$ excess(w) + cap
\EndProcedure
\State
\Procedure{Relabel}{$v$}
\For {all $e = \{v,w\} \in $ edge list of v}
	\State newd $\gets$ min \{ label(w) + 1, newd\}
\EndFor
\If {newd $>$ label(v)}
	\State label(v) $\gets$ newd
\EndIf
\EndProcedure
\end{algorithmic}
\end{algorithm}

\begin{algorithm}
\caption{Global Relabeling}
\label{alg:globalrelabel}
\begin{algorithmic}
\Procedure{GlobalRelabel}{}
	\If{currentlevel = 1}
    	\For {all vertices $v$ such that $e = \{v,s\}$ can carry flow}
        	\State get lock on $v$
        	\State wave($v$) $\gets$ global-wave-number
            \State label($v$) $\gets$ currentlevel
            \State add $v$ to processors global-relabel-queue[0]
        \EndFor
        \State set flag to indicate processor advanced frontier
    \Else
    	\State queuenumber $\gets$ mod(currentlevel,2);
    	\For {all vertices $v$ in processors global-relabel-queue[queuenumber]}
        	\State lock v if lockable
        	\For {all vertices $w$ such that $e = \{w,v\}$ can carry flow}
            \State try to get lock on $w$ if lockable
            \If {lock unsuccessful}
            	\State mark current edge position and push vertex $v$ back to global-relabel-queue[queuenumber]
                \State unlock v
                \State continue
            \EndIf
        	\State wave($w$) $\gets$ global-wave-number
            \State label($w$) $\gets$ currentlevel
            \State add $w$ to global-relabel-queue[mod(queuenumber+1,2)]
            \State unlock w
            \EndFor
        \EndFor
    \EndIf
\EndProcedure
\end{algorithmic}
\end{algorithm}

\pagestyle{plain}

\subsection{Implementation Details}
The details of the implementation are shown in detail in Algorithms ~\ref{alg:termination}, ~\ref{alg:vertexprocess} and ~\ref{alg:globalrelabel}. We implemented two versions of the algorithm one without implicit addressing and one with implicit addressing. For the first version, a "Edge" structure was used which stored pointer information to its mate edge, the node pointed by the edge and the next edge in the vertex's edge list apart from the edges residual capacity. Each vertex has an associated linked list of edges that begin from this vertex. In the version with implicit addressing scheme, the Edge structure was replaced with contiguous blocks of (residual) capacities. Each vertex has an associated vertex index. Depending on the edge interval of the proper order graph, each vertex has a fixed number of connections in any of the allowed directions (up, down, left, right, front, back and any combination of left/right/front/back with up and down). The starting edge for any vertex can be obtained by multiplying the total number of edges per a single vertex by the vertex's index. Vertices on the external boundaries are not treated specially, instead the non-existing edges have initial capacities set to zero so that they are not processed. In order to improve the cache performance, we ordered the vertices and edges such that the vertices that belong to a segment occur in consecutive memory locations and is therefore minimal false sharing across processors. Since a position of an edge in a vertex's edge block is direction based, an edge's mate can be found using the current edge's offset from the vertex's first edge. Since calculating these offsets could be an expensive operation, the offsets for every row/slice position and possible direction are cached in a 3D array.

The parallelism was achieved using boost threads. Once the graph was built, the user calls the maxflow procedure by passing the number of threads/segments to be created. The $s-t$ graph is then initialized and segmented into the required number of segments, with each thread assigned to process a particular segment. It is to be noted that the number of segments might be greater than the number of processors in the machine in which case, there might be contention for the computational resources. An additional  thread is also created which tracks the number of discharge operations currently issued and starts a new global relabel wave if the number of discharge operations exceeds a factor of the total number of nodes in the graph. This thread also helps in making sure that the global relabeling is done in a level synchronized manner. Cherkassy in ~\cite{Cherkassy:1995:IPM:645586.659457} suggests a global relabel factor of 1-2. We used the same range for our tests. In order to avoid unnecessary locks to update the discharge counter, the discharge counter was updated by each process only once after a fixed number of operations. In our tests, we found that updating the discharge operation at each thread by increments of 1000 vastly improved the speed of the algorithm.

Locking mechanisms are required only for the "shared" and "lockable" vertices that are between two adjacent segments. When flow is being pushed from a local non-lockable vertex $v$ to a lockable vertex $w$, the lockable vertex needs to be locked before pushing the flow. However, if the lockable vertex belonged to a different segment, then there is a possibility of a dead lock. Therefore in such cases, if the lock could not be obtained immediately, $v$ is pushed back to the FIFO queue and is unlocked. A similar strategy is used when pushing flow from a lockable vertex onto a non lockable vertex or between two lockable vertices. No locking is needed when the push operation is between two vertices that are interior to the segment. When relabeling, Anderson and Setubal in ~\cite{Anderson:1995:PIP:213123.213125} showed that is is sufficient to lock just the vertex being relabeled. In our implementation, we lock the vertex only if the vertex is lockable.

Both the process queues and the global queues are FIFO queues implemented as doubly linked list with the pointers to the next and previous elements in the queue being contained withing the Node data structure. Any update to the queues were carried out only after getting a lock on the queues. This is because the push and pop operations on a queue belonging to a segment, can also called by the adjacent segment adding active vertices to this segment. It is to be noted that a queue cannot be marked as empty as soon as the last element is popped out of it. This is because, the last element, though not part of the queue, might still be processed, and this vertex can, through push operations, make other vertices active. It is therefore imperative to wait for the last vertex to complete its discharge operations before marking the queues empty or otherwise, the program may lead to incorrect termination. The algorithm was implemented in C++ with boost libraries. The API was designed to be similar to the existing max-flow library used with Seg3D for easier integration.

\subsection{Results and Scalability Analysis}

\begin{figure}[ht]
\centering
\includegraphics[width=0.9\textwidth]{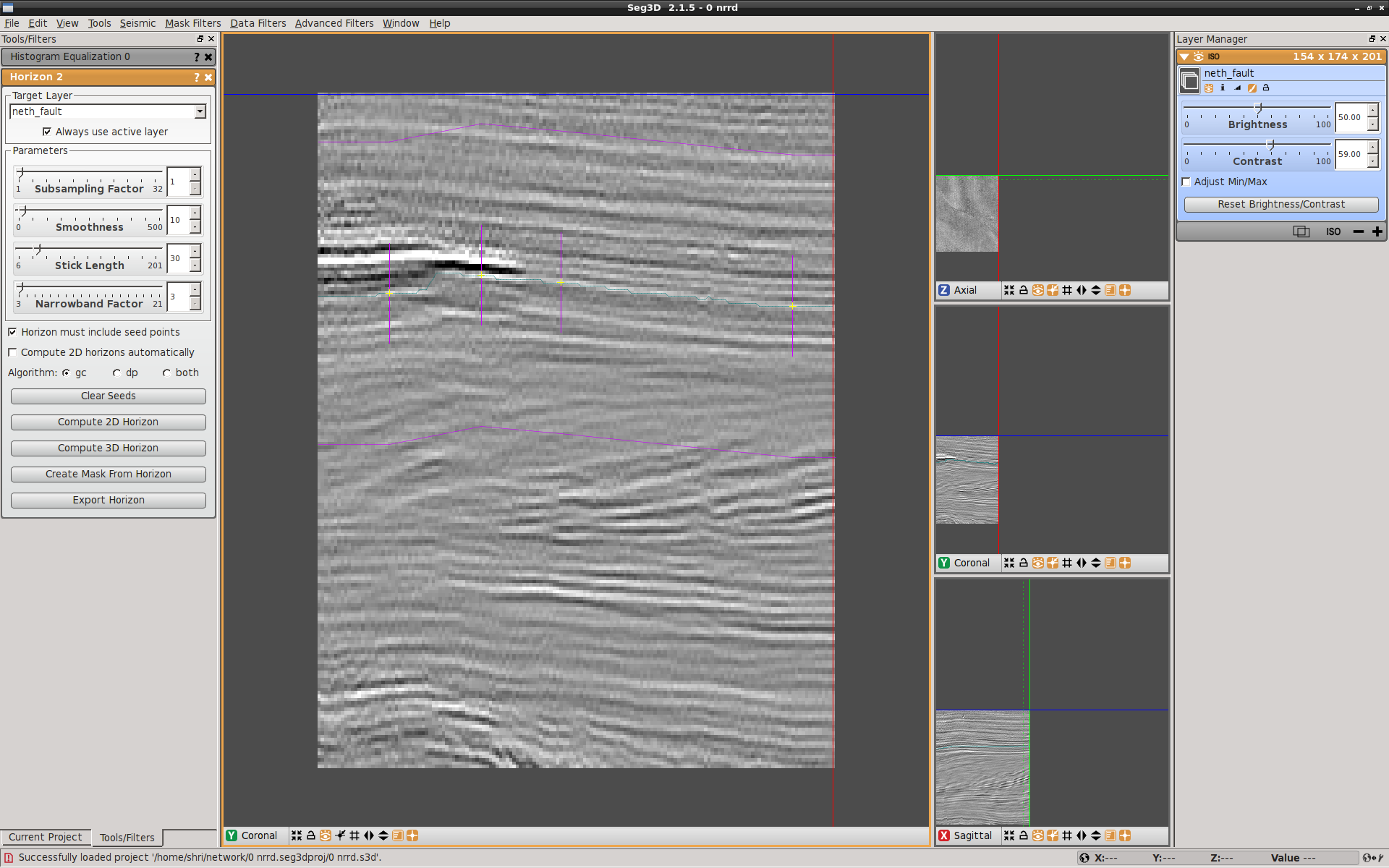}
\caption{\small{Figure shows the user interface for selecting the landmark points. The narrowband the stick length is marked by the vertical pink lines, whereas the narrow band factor affects the number of rows shown in the final image shown by the two horizontal pink lines.}}
\label{fig:narrowband}
\end{figure}

We ran a number of experiments predominantly on two seismic data volumes of different sizes as shown in Table ~\ref{table:imagesizes}. The segmentation problem here is finding the horizon surface in the seismic data. The user specifies a set of landmark points using a GUI integrated as part of Seg3D. The proper ordered mesh is then constructed around these landmark points and the $s-t$ mincut algorithm is run on the proper ordered graph in order to get the surface segmentation. In these experiments, the number of nodes in the proper-ordered graph depends on the narrow band factor and the stick length factors specified by the user, which limits the region of interest around these landmark points, and therefore limits the size of the structured graph on which the algorithm is actually run. The stick length denotes the number of pixels above and below the landmark points that need to be included which is then multiplied by the narrow band factor to get the number of rows in the structured proper ordered graph. The figure ~\ref{fig:narrowband} shows an example slice in an seismic image volume with the landmark points and the narrowband factors marked. In the experiments, we found that the current algorithm scales much better than the parallel BK-algorithm proposed by <cite adaptive>. All these experiments was run on a  shared computational machine with 80 Intel(R) Xeon(R) CPU E-7-4870 2.40GHz processors and 750 GB of RAM space.
\begin{table}
\centering
    \begin{tabular}{|c|c|}
        \hline
        Image & Size\\
        \hline
        Volume 1 & 551x426x426   \\
        Volume 2 & 1001x376x1375 \\
        \hline
    \end{tabular}
    \caption{\small{Table showing the test volume sizes.}}
    \label{table:imagesizes}
\end{table}

\begin{figure}[ht]
\centering
\includegraphics[width=0.9\textwidth]{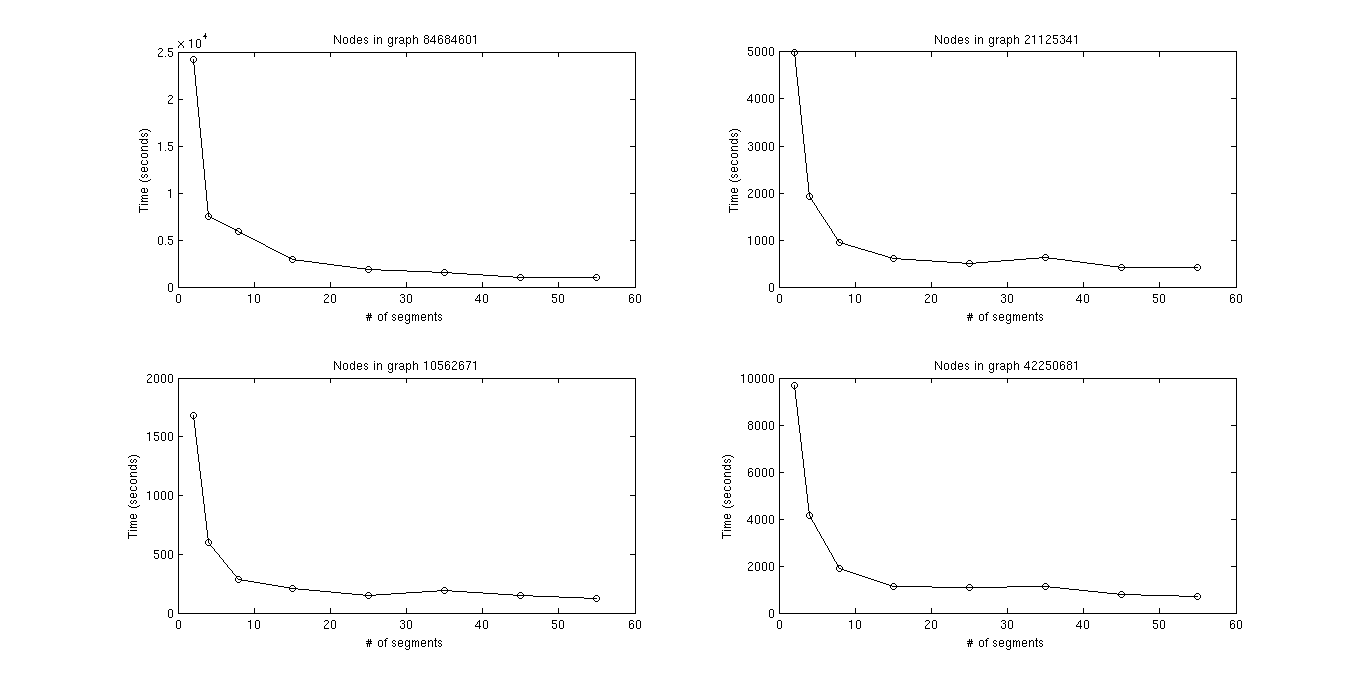}
\caption{\small{Plot of time against the number of segments(threads) shown for graphs with varying number of nodes}}
\label{fig:normalscaleplot}
\end{figure}

\begin{figure}[ht]
\centering
\includegraphics[width=0.9\textwidth]{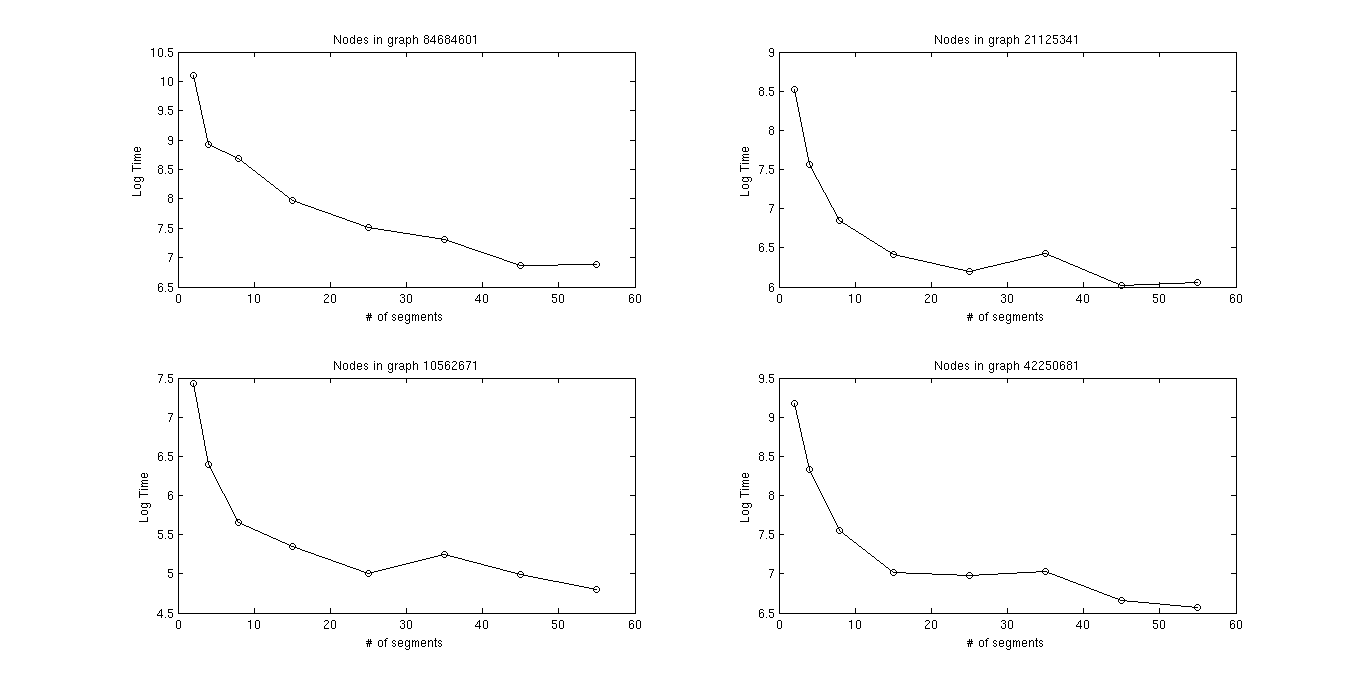}
\caption{\small{Plot of log time against the number of segments(threads) shown for graphs with varying number of nodes} The top left and the bottom right are volumes }
\label{fig:logscaleplot}
\end{figure}

The figure ~\ref{fig:normalscaleplot} shows the change in running times against the number of parallel threads spawned. The figure ~\ref{fig:logscaleplot} shows the change in log time against the number of segments. From the results, we are able to see that the algorithm scales well on graphs with a large number of nodes in it. For graphs of moderate size (10-20 million nodes) we got an almost linear decrease in running time on using 2-25 threads, and a plateau for any change in the number of threads beyond 25. For the larger volume (40-80 million nodes) we were able to see much better performance scaling. For instance in the tests run the test volume 1, we were able to see that the running time almost halved on doubling the number of processors. The amount of work that each thread has impacts the scaling factor of the algorithm. For relatively small graph volumes, the amount of work done locally against the work done on shared vertices decreases with increasing number of segments. This means that there is a point in the algorithm where the amount of performance improvement plateaus. The location of the plateau depends on the size of the volume as could be seen in the figure ~\ref{fig:normalscaleplot}.

\begin{figure}[ht]
\centering
\includegraphics[width=0.9\textwidth]{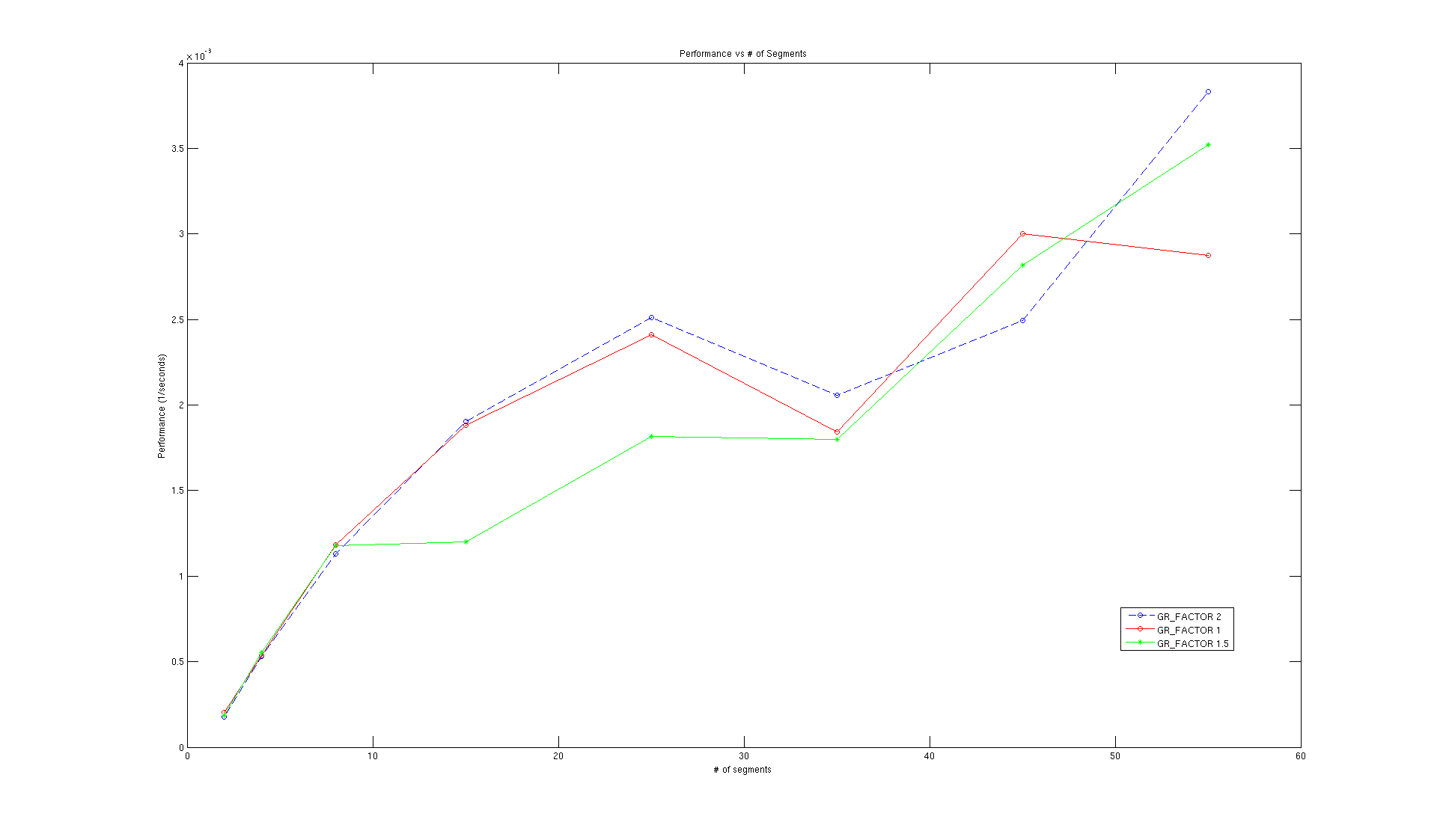}
\caption{\small{Plot of performance ($\frac{1}{time}$) against the number of segments(threads) shown for different global relabeling factors for test volume 2 with 21125341 nodes}}
\label{fig:grfactor}
\end{figure}

For the global relabeling factor, we found that the use of 2 (x number of nodes) for smaller graphs and 1 (x number of nodes) for larger graphs to give the optimal solutions. The choice of the global relabeling factor is one of empirical nature and the changes in the global relabeling factor within the range of 1-2 did not affect the performance significantly as shown in figure ~\ref{fig:grfactor}

\begin{figure}[ht]
\centering
\includegraphics[width=0.7\textwidth]{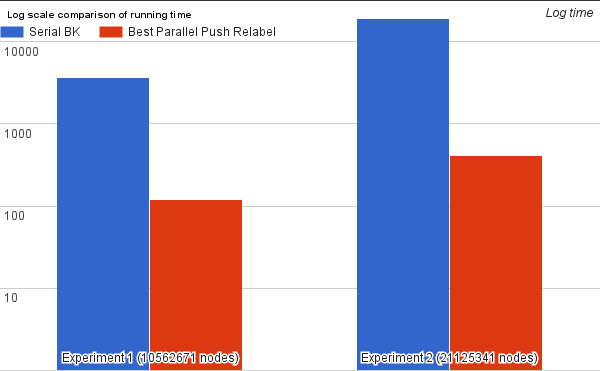}
\caption{\small{Comparison of serial running time (BK) vs best parallel Push Relabel time (log scale)}}
\label{fig:serialparallel}
\end{figure}

Figure \ref{fig:serialparallel} shows the time comparison of the running time for the serial BK and best parallel push relabel running times for two different experiments with 11 million and 21 million nodes. It is to be noted that the current serial version of the BK algorithm failed to work for graphs with a larger number of nodes due to the high memory requirements of the unstructured graph structures. It is also to be noted that the results compare the timings of the parallel version of one algorithm with the serial version of an existing but different algorithms.

\begin{figure}[ht]
\centering
\includegraphics[width=0.7\textwidth]{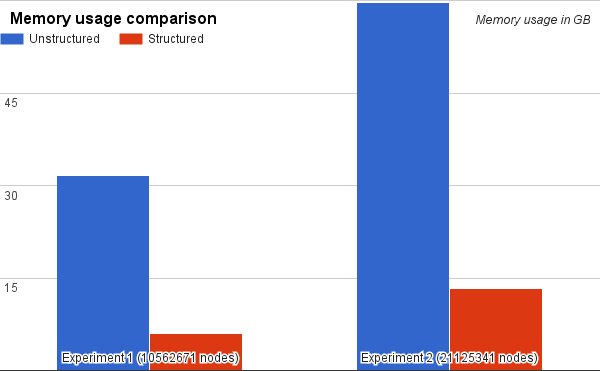}
\caption{\small{Comparison of memory usage for the structured and unstructured versions of the parallel push relabel algorithm}}
\label{fig:memorycompare}
\end{figure}

Finally the figure in \ref{fig:memorycompare} shows the memory consumption comparison between the structured and the unstructured versions of the parallel push relabel version. For the unstructured version, each half-edge took 32 bytes of memory and each node needed 128bytes of data. For any proper order graph, the number of edges in the graph can outnumber the number of nodes by a factor of 10-100. For an edge interval of 10, each node has 88 halfedges associated with it. For the structured graph, each half edge required exactly 4 bytes of data and each node required 128 bytes. The structured version also required a cache memory that occupied a space proportional to the number of rows and slices in the volume.

\pagestyle{plain}

\clearpage
\section{Conclusion}
\label{sec:conclusion}
\par
This project has extended the parallel push relabel algorithm first proposed by Anderson and Setubal and extended it to perform scale well for a particular class of structured meshes and also makes the implementation "cache aware". The algorithm both scales well and is memory efficient to be used in practical image segmentation algorithms. While this implementation provided good performance results, in future study, it will be interesting to study the impact of Highest Label vertex processing and Gap relabeling. Both these heuristics have shown to provide good results in the serial setting. In the parallel setting however, the gap relabeling strategy can be tricky to implement. This is because, the current partitioning strategy does not allow one processor to know about the contents of the vertex queue in another processor and therefore gap relabeling will require and additional level of global information tracking which could potentially be a hindrance to parallel performance.

\pagestyle{plain}

\renewcommand{\refname}{References}
\bibliography{report}

\end{document}